%% file: resources_scaling.tex
\documentclass{aidb}

\usepackage{bbm}

\newcommand{\todo}[1]{}
\renewcommand{\todo}[1]{{\color{red} TODO: {#1}}}

\newcommand{\header}[1]{\vspace{1mm}\noindent\textbf{#1}.}

\usepackage{tikz}
\usetikzlibrary{fit,positioning}

\usepackage[framemethod=tikz]{mdframed}
\usepackage{xcolor}
\definecolor{mycolor}{rgb}{0,0.0,0.0}

\input{macros.tex}

\begin{document}

\title{A simple and effective predictive resource scaling heuristic for large-scale cloud applications}

\numberofauthors{5}
\author{
\alignauthor Valentin~Flunkert\\
	\affaddr{Amazon Research}\\
	\email{flunkert@amazon.com}\\
\alignauthor Quentin~Rebjock\\
	\affaddr{EPFL}\\
	\email{quentin.rebjock@epfl.ch}\\
\alignauthor Joel~Castellon\\
	\affaddr{Amazon Research}\\
	\email{jolcast@amazon.com}\\
	\and
\alignauthor Laurent~Callot\\
	\affaddr{Amazon Research}\\
	\email{lcallot@amazon.com}\\
\alignauthor Tim~Januschowski\\
	\affaddr{Amazon Research}\\
	\email{tjnsch@amazon.com}\\
}
\date{\today}
\maketitle

\begin{abstract}
We propose a simple yet effective policy for the predictive auto-scaling of horizontally scalable applications 
running in cloud environments, where compute 
resources can only be added with a delay, and where the deployment throughput is limited.
Our policy uses a probabilistic forecast of the workload to make scaling decisions dependent on the risk aversion of the application owner.
We show in our experiments using real-world and synthetic data that this policy 
compares favorably to mathematically more sophisticated approaches as well as to simple benchmark policies.
\end{abstract}

\input{main.tex}

\bibliographystyle{ieeetr}
\bibliography{papers}
\end{document}

%% file: macros.tex
\usepackage{mathtools}
\usepackage{dsfont}
\usepackage[T1]{fontenc}
\usepackage{color}
\usepackage{graphicx}
\usepackage{wrapfig}
\usepackage{xparse}
\usepackage{enumitem}
\usepackage{multirow}
\usepackage{array}
\usepackage{blkarray}
\graphicspath{{figs/}}  

\usepackage{algorithm}
\usepackage{algpseudocode}
\usepackage{booktabs}
\usepackage{verbatim}
\usepackage{subcaption}

\DeclarePairedDelimiterX{\lin}[2]{\langle}{\rangle}{#1, #2}
\DeclarePairedDelimiterX{\abs}[1]{\lvert}{\rvert}{#1}
\DeclarePairedDelimiterX{\norm}[1]{\lVert}{\rVert}{#1}
\DeclarePairedDelimiterX{\cbr}[1]{\{}{\}}{#1} 
\DeclarePairedDelimiterX{\rbr}[1]{(}{)}{#1} 
\DeclarePairedDelimiterX{\sbr}[1]{[}{]}{#1} 

  \DeclareMathOperator{\sgn}{sign}
  \makeatletter
  \def\sign{\@ifnextchar*{\@sgnargscaled}{\@ifnextchar[{\sgnargscaleas}{\@ifnextchar{\bgroup}{\@sgnarg}{\sgn} }}}
  \def\@sgnarg#1{\sgn\rbr{#1}}
  \def\@sgnargscaled#1{\sgn\rbr*{#1}}
  \def\@sgnargscaleas[#1]#2{\sgn\rbr[#1]{#2}}
  \makeatother

  \providecommand{\tt}{\mathbf{t}}

  \providecommand{\mZ}{\mathbf{Z}}

  \providecommand{\posreq}{q}
  \providecommand{\negreq}{f}
  \providecommand{\delay}{\tau}

  \providecommand{\hdelay}{\hat{\tau}}



%% file: main.tex

\section{Introduction} \label{sec:introduction}

Modern cloud computing providers automate the scaling of hardware resources to meet
the demand of the hosted application. The key consideration when designing scaling policies is
the trade-off between reducing costs by minimizing the allocated resources 
and satisfying customers by supplying sufficient resources for the application to run nominally. 
Auto-scaling is a well-studied topic for which comprehensive
surveys and reviews are available \cite{Qu2018Survey, makridakis2018m4}.

We consider complex, large-scale applications with hundreds or thousands of
servers, for which reactive auto-scaling policies often exhibit shortcomings
stemming from unrealistic assumptions. 
For example, while releasing hosts is usually fast, there
are limits to how quickly the cloud provider can fulfill requests to add more
instances to a fleet, and in particular for large volumes.
Reactive scaling assuming instantaneous resource addition will then lag behind the true demand and
the fleet may be at risk of being under-provisioned before a traffic peak and over-provisioned
afterwards.
Predictive scaling can circumvent these limitations by using a
forecast of the application's workload to make scaling decisions ahead of time while
taking throughput constraints into account.
Using a probabilistic forecast, we can design policies that are optimal for the
level of risk aversion of the application owner, in the sense that any quantile
of the random workload can be estimated.
This approach is particularly well suited for applications with
strongly seasonal (daily) traffic patterns and for large fleet sizes, for
which the time to scale the fleet up is significant.

Our contributions to predictive auto-scaling for cloud applications are as
follows.
\textit{\textbf{(i)}} We develop an approach for analyzing and evaluating auto
scaling policies for cloud applications. Our approach includes a risk aware cost
function and a realistic model for the scaling behavior of the application
taking into account real world constraints on throughput and latency.
\textit{\textbf{(ii)}} We show how \emph{probabilistic} forecasts naturally lend
themselves to satisfy the risk aversion of the owner of the application.
\textit{\textbf{(iii)}} We show through experiments that a simple heuristic becomes
optimal in the case of high risk aversion which is the practically most relevant
case.

Similar heuristics are used to auto-scale over 40\,000 of Amazon's internal
auto-scaling groups, as well as external applications making use of AWS
Auto-Scaling\footnote{https://aws.amazon.com/autoscaling/}.

We discuss related work in Sec.~\ref{sec:related-work}. We then introduce and
formalize the predictive auto-scaling problem and state our modeling assumptions
in Sec.~\ref{sec:problem}. We present the scaling policies and evalute them
empirically in Sec.~\ref{sec:scaling_policies} using synthetic and real-world
data.

\section{Related Work} \label{sec:related-work}
Scaling resources in cloud environments is a mature and active area of research,
see e.g.~\cite{Qu2018Survey, makridakis2018m4} 
for reviews.
Predictive scaling has been studied alongside reactive scaling
(\cite{Wu2019Anna} is a recent example in the context of storage systems) and
products such as AWS Auto-Scaling now offer both predictive and reactive scaling
techniques. Hybrid methods taking advantage of the strengths of both schemes
have also been studied~\cite{coordinated}.

Approaches that rely on \emph{forecasting} techniques
(like~\cite{Messias2016,Kirchhoff19,Lang2016,Ma2018,taft2018}) are the most
closely related to our work
but state of the art probabilistic forecasts are rarely considered.
They are, however, key for optimal decision
making~\cite{Faloutsos2018,Faloutsos2019}.
This is well-known in the supply chain literature~\cite{simchi13logistics}.
In many settings, neural network architectures for forecasting have been shown
to deliver superior predictive accuracy than traditional
techniques~\cite{smyl2018m4,deepAR,rangapuram2018deep}. They are used in AWS
auto-scaling for this reason.

An exception is~\cite{Borkowski2019} which relies on a probabilistic time series
model to infer the true state of the application (but not to forecast workload).
The discovery of this state allows them to handle the cost of scaling
operations, a problem which is more relevant in a streaming scenario than in our
scenario.
Their use of multiple metrics in a multivariate time series models is relevant
for the subject matter of this paper,
it should be explored in future work
(e.g., through using modern multivariate forecasting models~\cite{multivariate}).

\section{Problem statement} \label{sec:problem}

In this section we formalize the predictive scaling problem and discuss our modeling assumptions.
We consider a fleet of servers that run a horizontally scalable application.
Our goal is to scale the number of servers in an \emph{optimal} way to match the actual demand.

We assume a work-load estimation model to be known, outputting an approximation
of the demand (ideal number of hosts) for the application,
given an observable metric.\footnote{
  This is no small assumption. Depending on the application,
  modeling the work-load may require significant work and it is a research area
  in its own right e.g.,~\cite{Lang2015,SchaffnerJ13,Duggan2011,Mozafari2013,Ma2018,das2016automated}.
}
The selected target metric must not be derived from the scaling decisions for
the forecasting model to be an unbiased predictor of the demand.
The sum of the number of bytes received and sent by every server in the fleet is one such measure, which we found to be pertinent as it is strongly correlated with the
CPU and memory usage of the servers.

We denote this variable by $v_t$ and a simple linear workload model yields that
the number of desired hosts $z_t$ is proportional to the measured load $z_t = \xi v_t$.
In practice the function linking $v_t$ to $z_t$ may be more complex and
estimated with more sophisticated models.

\subsection{Probabilistic Forecast and cost function}\label{subsec:forecast_and_cost}
Given the workload representation $v_t$, a forecasting model can be trained to
generate estimates of the probability distribution for the future
  $P(v_{t+1}, v_{t+2}, \dots, v_{t+T} \mid \mbox{past})$.\footnote{
  We assume discrete time steps. Real cloud events and scaling actions occur in continuous time,  however a discretization in steps that
  match real scaling time-frames avoids requiring complex time series models
  such as~\cite{turkmen19}.
  Delays for the host provider are a few minutes for releasing or a few dozens
  of minutes to acquire new hosts.
  Therefore, a discretization into minute steps seems a non-restrictive assumption.
}
In practice, workload trends are dominated by daily and weekly patterns. 

In order to make optimal scaling decisions we need to define the cost function
that describes the cost for over-provisioning and under-provisioning the fleet.
We assume that the cost is composed of two terms.
The first one is hardware cost, which
scales linearly with the number of provisioned hosts.
The second term captures the service's performance. This may be, for instance, loss of revenue or customers
caused by the increase in latency when the service is under-provisioned.

The latter part is harder to quantify and measure. We make the following
assumption: at each time step $t$ there is a critical capacity $r^\star_t$ such
that the service will perform optimally if the provisioned capacity $r_t$ is
above $r^\star_t$. 
Adding more capacity than $r^\star_t$ does not improve the performance of the
service while increasing the operational cost. If $r_t < r^\star_t$, the
application's performance deteriorates. The cost associated with this is
proportional to $r^\star_t - r_t$, typically with a large proportionality constant. 
In practice $r^\star_t$ is never observed, but $z_t$ should approximate it
closely if the workload model is good.

This translates into an asymmetric cost function which is, up to a constant
factor, given by the $\alpha$-quantile loss function (sometimes also called the pinball loss, 
see~\cite{gneiting2007probabilistic} for a thorough discussion) described in
Equation~\eqref{eq:pinball}.
\begin{equation}\label{eq:pinball}
  \Lambda_\alpha(r_t, z_t) = \begin{cases}
    (1 - \alpha) \cdot (r_t - z_t) & \mbox{if }r_t > z_t \;,\\
    \alpha \cdot (z_t - r_t) & \mbox{otherwise.}
  \end{cases}
\end{equation}
The quantile $\alpha\in (0,1)$ corresponds to the risk aversion factor.
Note that $z_t$ is a random variable as the forecast of $v_t$ is probabilistic.
The expected cost $\mathbb{E}[ \Lambda(r_t, z_t) ]$ is minimized when $r_t =
q_\alpha(z_t)$, i.e., the $\alpha$-quantile of $z_t$.

\subsection{Model for cloud provider behavior}

A central entity in our problem statement is the cloud or host provider that
provisions the instances for the fleet (e.g.\ the AWS EC2 service).
The application owner can request new instances from this provider or release
superfluous instances.

\begin{figure}[htbp]
	\centering
	  \begin{tikzpicture}[scale=0.4]
	  \node[draw, rounded corners=4pt,align=center] at (4,3) {$R$ outstanding requests};
	  \draw [rounded corners=4pt] (0,0) rectangle ++ (8,1);
	  \draw[fill=lightgray] (0.5,0.5) circle (0.3cm);
	  \draw[fill=lightgray] (1.5,0.5) circle (0.3cm);
	  \draw[fill=white] (2.5,0.5) circle (0.3cm);
	  \draw[fill=lightgray] (3.5,0.5) circle (0.3cm);
	  \draw[fill=white] (4.5,0.5) circle (0.3cm);
	  \draw[fill=white] (5.5,0.5) circle (0.3cm);
	  \draw[fill=white] (6.5,0.5) circle (0.3cm);
	  \draw[fill=lightgray] (7.5,0.5) circle (0.3cm);
	  \draw [->,line width=0.8] (4,2.3) --  node [text width=4cm,midway,right] {wait for free slot} (4,1.2);
	  \draw [->,line width=0.8] (8,0.5) --  node [] {} (10.8,0.5);
	  \node[draw, rounded corners=4pt,align=center] at (15,0.5) {alive after time $\tau \sim P(\tau)$};
	  \end{tikzpicture}
	 \caption{\label{fig:providermodel}Process for provisioning new hosts.}
\end{figure}
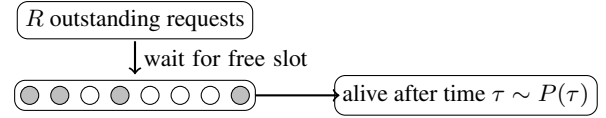

While the exact behavior of the host provider may be complicated, we make the
following simplifying assumptions:
\textit{(i)} the provider has an infinite reservoir of instances, 
\textit{(ii)} releasing an instance back to the provider is instantaneous and
\textit{(iii)} requests for new instances are handled as depicted in
Figure~\ref{fig:providermodel}.
That is, the provider has a fixed number of slots for provisioning,
which can be empty (white) or occupied (gray). 
The number of hosts requested by the user are summed to $R \ge 0$.
When slots are empty they are immediately filled and $R$ is decreased.
When a slot is filled, the corresponding instance is being provisioned and it
takes a (random) time $\tau$ until the host is available.
At this point the slot becomes empty again.
For most applications, these assumptions were found not to be overly
restrictive.

\begin{figure}[htpb]
  \centering
  \includegraphics[width=0.7\linewidth, height=3cm]{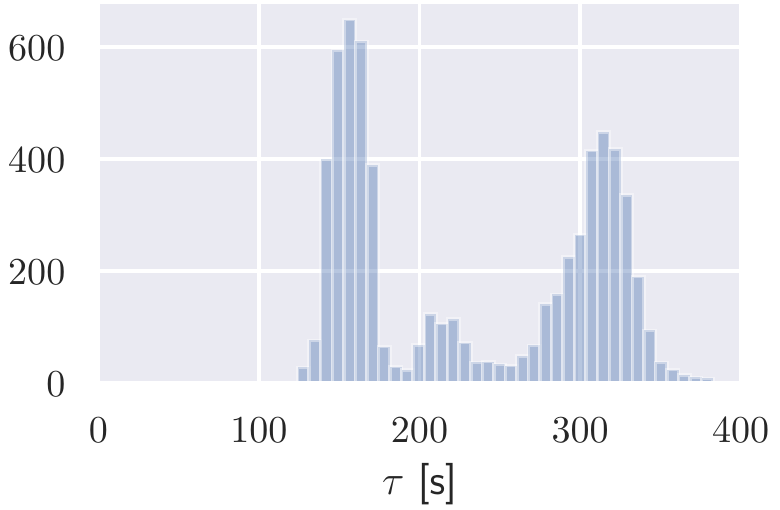}
  \\[-3.5ex]\caption{
    \label{fig:providerdelay}Histogram of $P(\tau)$ from individual
    scaling events of a real application (count on vertical axis).
  }
\end{figure}

We denote by $\tau$ the delay between asking for additional resources and those 
resources becoming available.
Its distribution $P(\tau)$ depends not only on how quickly the
provider can spin up a new instance, but also on the time it takes
to install and deploy the necessary packages, the data (e.g.\ cache), and to start
the application.
Typical values for real-world applications are between
a few minutes and half an hour.
This can vary significantly with the application and the deployment approach.
For these reasons, the resource type can exhibit a complex structure as
illustrated by the histogram of delays in Figure~\ref{fig:providerdelay}.

When the fleet size (and the corresponding scaling amounts) is large, the
limited number of slots in Figure~\ref{fig:providermodel} translates into an
effective rate limit $\rho$ for the number of hosts that can be added per unit
of time.
Our assumptions are simplications: the delays and the throughput $\rho$ are likely to be coupled, to vary over time, and to depend on the current and past request values. However, we found
these assumptions to approximate the real world behavior well.

\subsection{Scaling Cost Model}
Next, we combine the aforementioned models to calculate the \emph{expected cost} of the scaling policies.

Let us fix a time horizon $n > 0$ and consider a probabilistic forecast of length
$n$ generated at time $0$.
Using Equation~\eqref{eq:pinball}, the expected cost for a sequence of future resource values $r = (r_1, \dots, r_n)$ is
\begin{align}
  \mathcal{L}_\alpha(r) &= \sum_{i=1}^{n} \int \Lambda_\alpha(r_i, z_i) P(z_1, \dots, z_n)  \,d z_1 \dots d z_{n} \nonumber\\
                       &= \sum_{i=1}^{n} \mathbb{E}_{z_i\sim P(z_i)} \; \Lambda_\alpha(r_i, z_i) \;.
\end{align}
We denote the actual cost by $\mathcal{L}_\alpha^\star(r) = \sum_{i=1}^{n}
\Lambda_\alpha(r_i, r^\star_i)$, which is dependent on the true demand $r^\star$
and cannot be directly minimized.
If the workload model is adequate (that is, $z$ is a good estimate of
$r^\star$), minimizing $\mathcal{L}_\alpha(r)$ should provide a reasonable
resources pattern for the loss $\mathcal{L}_\alpha^\star(r)$. The choice of
$\alpha$ depends on the relative costs of the service being under- or
over-provisioned as discussed in Section~\ref{subsec:forecast_and_cost}.

The optimal solution of $\mathcal{L}_\alpha(r)$
is $r = \left(q_\alpha(z_1), \dots, q_\alpha(z_n)\right)$ where $q_\alpha(z_i)$
is the $\alpha$-quantile of $z_i$.
However, we cannot act on $r_t$ directly; instead, at each time step $t \in \mZ$
we are able to request $\posreq_t \in \mathbb{N}$ and release $\negreq_t \in
\mathbb{N}$ hosts.
As pointed out above, releases are instantaneous while every single requested
instance will arrive $\tau$ time steps later.
As we have a forecast of the demand for the next $n$ time steps,
we want to optimally pick the values $\posreq_0, \dots, \posreq_n$ and
$\negreq_0, \dots, \negreq_n$,
given the past requests $\posreq_{t' < 0}$ and $\negreq_{t' < 0}$ and the past
resources values $r_{t' \le 0}$.
The larger the forecast horizon $n$, the earlier scaling decisions can be taken.
But forecasting accuracy decreases as the horizon increases, so the choice of
$n$ is a trade-off that depends on the periodicity of the time-series and on the
effectiveness of the forecasting model.

\header{Resources estimation} Denote by $r_{-\infty}$ the host count at the start of
the auto-scaling process, before taking any decision.
If we are able to estimate accurately the host provider's behavior, we can
compute for each $t$ an approximation $\hat r_t$ of $r_t$ as a function of $\posreq_{t' \le t}$ and $\negreq_{t' \le t}$.
Let $\hat\delay$ be an estimation of the true host provider random delay
for acquiring new instances.
Then the random resource estimation at time $t$ can be written as
\begin{equation}
\hat{r}_t(q, f) = r_{-\infty} + \sum_{i \le t} \bigg( \sum_{j = 1}^{\posreq_i}
  \mathbbm{1}\{i + \hat\delay_{i, j} \le t\} - \negreq_i \bigg)
\end{equation}
where $(\hat\delay_{i, j})_{i, j \in \mZ} \overset{i.i.d.}{\sim} \hat\delay$
is a collection of random variables representing the delays of the potential
positive requests.
When the number of hosts is large, it may be approximated in
expectation in the following way
\begin{equation}
  \hat{r}_t(q, f) = r_{-\infty} + \sum_{i \le t} \Big( \posreq_i \cdot
  \mathbbm{P}\{i + \hat\delay \le t\} - \negreq_i \Big)
\end{equation}
by using the fact that, in expectation, a request $q$ done at time $t$ will yield $q \cdot
\mathbbm{P}\{\hdelay = \delta\}$ hosts $\delta$ time steps later.

\header{Optimization Problem Formulation} In summary, we wish to solve the following optimization problem, 
given that $\posreq_{t' < 0}$ and $\negreq_{t' < 0}$ are fixed.
\begin{equation}\label{eq:optimization_problem}
\begin{aligned}
	& \underset{\substack{\posreq_0, \dots, \posreq_n, \\ \negreq_0, \dots, \negreq_n}} {\text{minimize}}
	& & \mathbb{E}_{z} \, \mathbb{E}_{\hat{r}(\posreq, \negreq)} \, \mathcal{L}\left( \hat{r}(\posreq, \negreq), z \right)  \\
	& \text{subject to}
	& & \posreq_i \le \hat\rho, \negreq_i \leq r_i, \posreq_i \in \mathbb{N}, \negreq_i \in \mathbb{N}
\end{aligned}
\end{equation}
where $\hat\rho$ is an estimation of the throughput $\rho$.
The objective function in Equation~\eqref{eq:optimization_problem} is convex
as the quantile loss is convex and $\hat{r}$ is linear.
The integrality constraints make the optimization challenging but, assuming a
large-enough number of hosts, a relaxation followed by a randomized rounding produce
near-optimal solutions.

The double expectation in Equation~\eqref{eq:optimization_problem} may be
approximated with Monte-Carlo averaging; it only requires to sample from $z$ and
from $\hat r(q, f)$ (the latter being done with samples of the delay $\hat\tau$).
From here, the optimization of Equation\eqref{eq:optimization_problem} results in minimizing
a convex piecewise-linear function. This problem is equivalent to solving a
linear-program, which can be done efficiently.
Once the objective is optimized, we can request the $k$ first values at the
right time and repeat the process $k$ time steps later in what amounts to 
lookahead optimization.

The variant of the auto-scaling problem that we consider in this paper is, to
the best of our knowledge, new to the literature.
Prior work either assumes the cloud to be fully elastic and resources to come
available immediately or with a delay (e.g., ~\cite{Wu2019Anna}),
but the constraints introduced by the limited throughput of the cloud provider have not yet been studied.

\section{Policies \& Experiments}\label{sec:scaling_policies}

In the following, we describe the scaling policies that we compare empirically.
Forecasts are made every hour, over two days, and with 5 minutes granularity.
Every forecast overrides the previous one as it is considered to be more accurate.
Whenever a new forecast is made, new decisions are taken for the next hour,
where requests and releases are planned with a granularity of 5 minutes.
The choice of these parameters depends greatly on the workload patterns that
must be addressed;
we found these to be adapted to the daily trends that are very common at AWS.

\header{Policies}
The following two policies are frequently used in industrial predictive
auto-scaling applications~\cite{Qu2018Survey}.
The \emph{maximum observed needed capacity} policy keeps the host count at the
maximum that was needed in a recent past (for example one day or one week).
That kind of policy is very conservative but commonly employed due to its robustness.
Another baseline, \emph{reacts periodically to the current need}.
That is, every 5 minutes, it estimates the desired host count $z = \xi v$ and requests
resources accordingly.
The estimation can be adjusted by some factor in order to be more or less conservative.
This is better known as \emph{reactive} scaling and usually performs well if the
resource needs do not vary too rapidly.

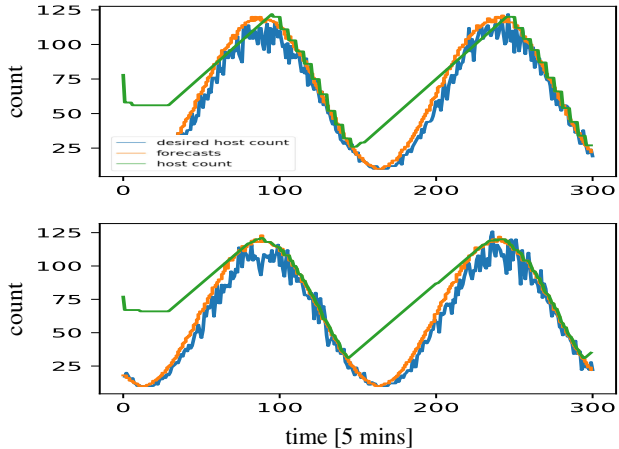
\begin{figure}[!h]
  \begin{subfigure}[t]{0.5\textwidth}
    \begin{tikzpicture}
      \centering
      \node (img1) {
        \includegraphics[width=\linewidth, height=3cm]{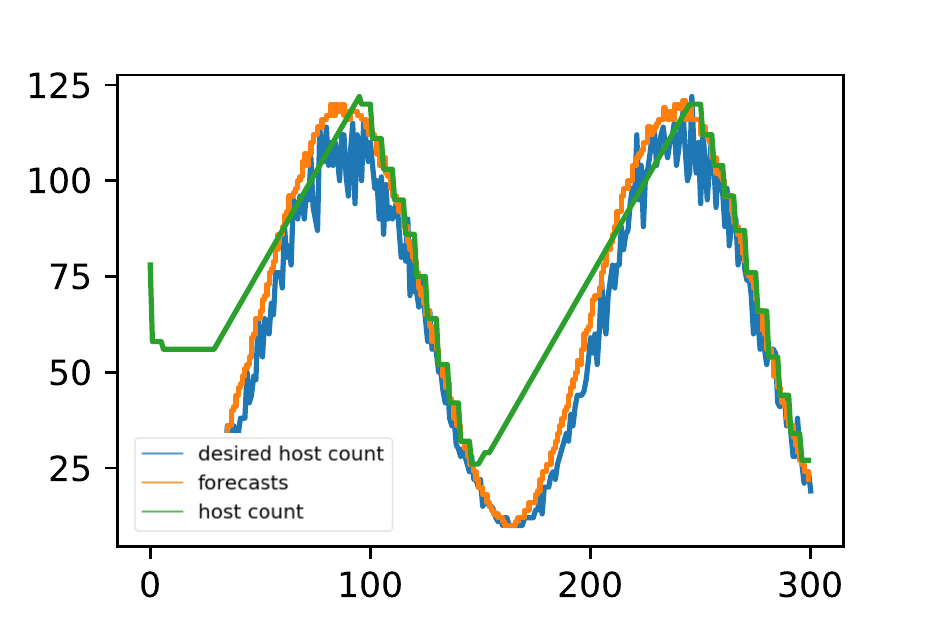}
      };
      \node[left=of img1, node distance=0cm, rotate=90, anchor=center,yshift=-1.1cm,font=\color{black}] {count};
    \end{tikzpicture}
  \end{subfigure}\\[-2.5ex]
  \begin{subfigure}[t]{0.5\textwidth}
    \begin{tikzpicture}
      \centering
      \node (img1) {
        \includegraphics[width=\linewidth, height=3cm]{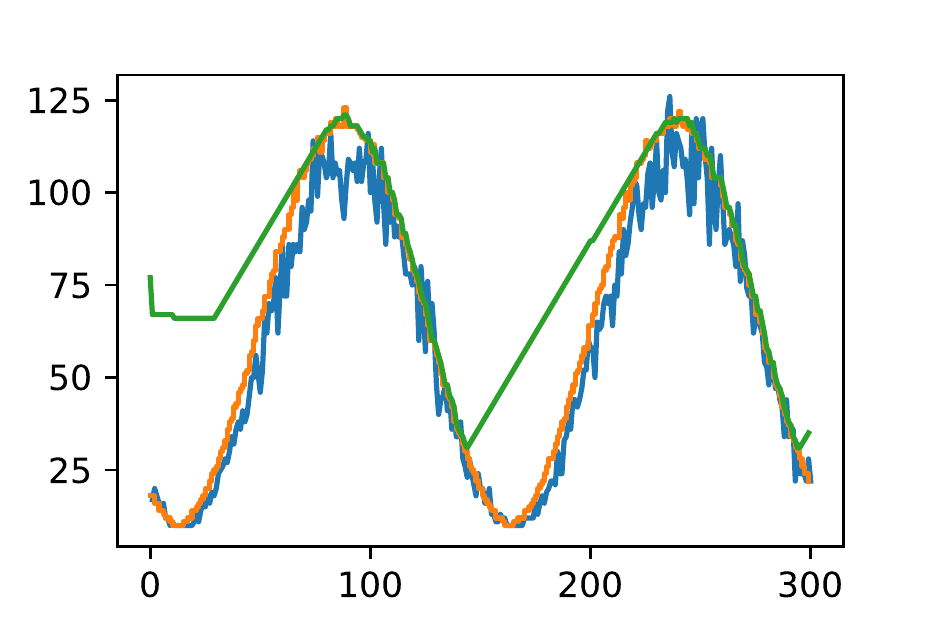}
      };
      \node[below=of img1, node distance=0cm, yshift=1.15cm,font=\color{black}] {time [5 mins]};
      \node[left=of img1, node distance=0cm, rotate=90, anchor=center,yshift=-1.1cm,font=\color{black}] {count};
    \end{tikzpicture}
  \end{subfigure}\\[-4ex]
  \caption{
    \label{fig:qshift}
      Host counts for optimization (top) and shifting (bottom) policies
      simulated on $\boldsymbol{D}_1$. The quantile $\alpha = 0.9$ of the
      forecasts cannot be always matched because of the limiting throughput
      $\rho$.
  }
\end{figure}

We introduce the simple \emph{forecast shifting} policy based on forecasts that
is described in Algorithm~\ref{alg:forecast_shifting}. It works in three steps:
\emph{(i)} computing the $\alpha$-quantile of the forecast $z$ and adjusting it
backward to take into account the limiting throughput $\rho$,
\emph{(ii)} splitting the positive and negative requests, and
\emph{(iii)} shifting the positive ones by the mean of the delay $\hat\tau$ to
call for them earlier.
\begin{algorithm}
    \caption{Forecast shifting}\label{alg:forecast_shifting}
    \begin{algorithmic}[1]
        \State \textbf{Input:} Probabilistic forecast $z = (z_1, \dots, z_n)$,
        guessed random delay $\hat{\tau}$ and throughput $\hat{\rho}$, current host count $r_0$
        \State $z'_t \gets \text{quantile}_\alpha(z_t) \; \forall t$ \algorithmiccomment{\emph{(i)}}
        \For{$t = n - 1,\dots, 1$}
            \State $z'_t := \max(z'_t, z'_{t + 1} - \hat{\rho})$
        \EndFor
        \State $\Delta_t \gets z'_t - z'_{t - 1}$ for $t = 2,\dots, n$ and $\Delta_1 = z'_1 - r_0$
        \algorithmiccomment{\emph{(ii)}}
        \State $q,\,f \gets \max(\Delta, 0),\,\min(\Delta, 0)$
        \State Shift $q$ in the past according to $\mathbb{E}[\hat\delay]$ \algorithmiccomment{\emph{(iii)}}
        \State Request $q$ and release $f$
    \end{algorithmic}
\end{algorithm}
Figure~\ref{fig:qshift} (bottom) illustrates the backward adjustment performed in step
\emph{(i)}. The forecasts (orange) must be compensated to take into account the
throughput $\rho$.
This algorithm can be adapted to take into account the full distributions of the
delay rather than only the expectation. However, all our experiments showed that
considering the full distribution was not leading to any loss improvement.

For large risk aversion the cost of underpredicting is much higher than over
prediction. Without constraints the $\alpha$-quantile is the optimal solution
for this asymmetric trade-off. In the limit $\alpha \to 1$ the optimal capacity
with constraints is equal to the $\alpha$-quantile whenever possible but never
below it. This is what the forecast shifting method achieves.

The \emph{optimal policy} is to solve the problem as posed in
Equation~\eqref{eq:optimization_problem} with standard mathematical programming
frameworks like SCIP~\cite{GleixnerBastubbeEifleretal.2018}.

\header{Datasets and predictors} In order to compare the 4 scaling methods
mentioned above, we run experiments in a simulated environment (using the
framework SimPy~\cite{simpy}) on one real-world ($\boldsymbol{R_1}$) and two
artificial ($\boldsymbol{D_1}$ and $\boldsymbol{D_2}$) datasets.
The real-world dataset $\boldsymbol{R_1}$ is composed of $1\,000$ randomly
selected Amazon auto-scaling groups over a period of 6 hours.
The datasets $\boldsymbol{D_1}$ (low noise) and $\boldsymbol{D_2}$ (high noise)
are generated with strong daily/weekly patterns, some linear trends, and
Gaussian noise. For all of them, a DeepAR~\cite{deepAR} forecasting model is
trained and makes new predictions every hour.

\begin{figure}[!h]
  \begin{subfigure}[t]{0.5\textwidth}
    \begin{tikzpicture}
	    \centering
      \node (img1) {
        \includegraphics[width=\linewidth, height=4cm]{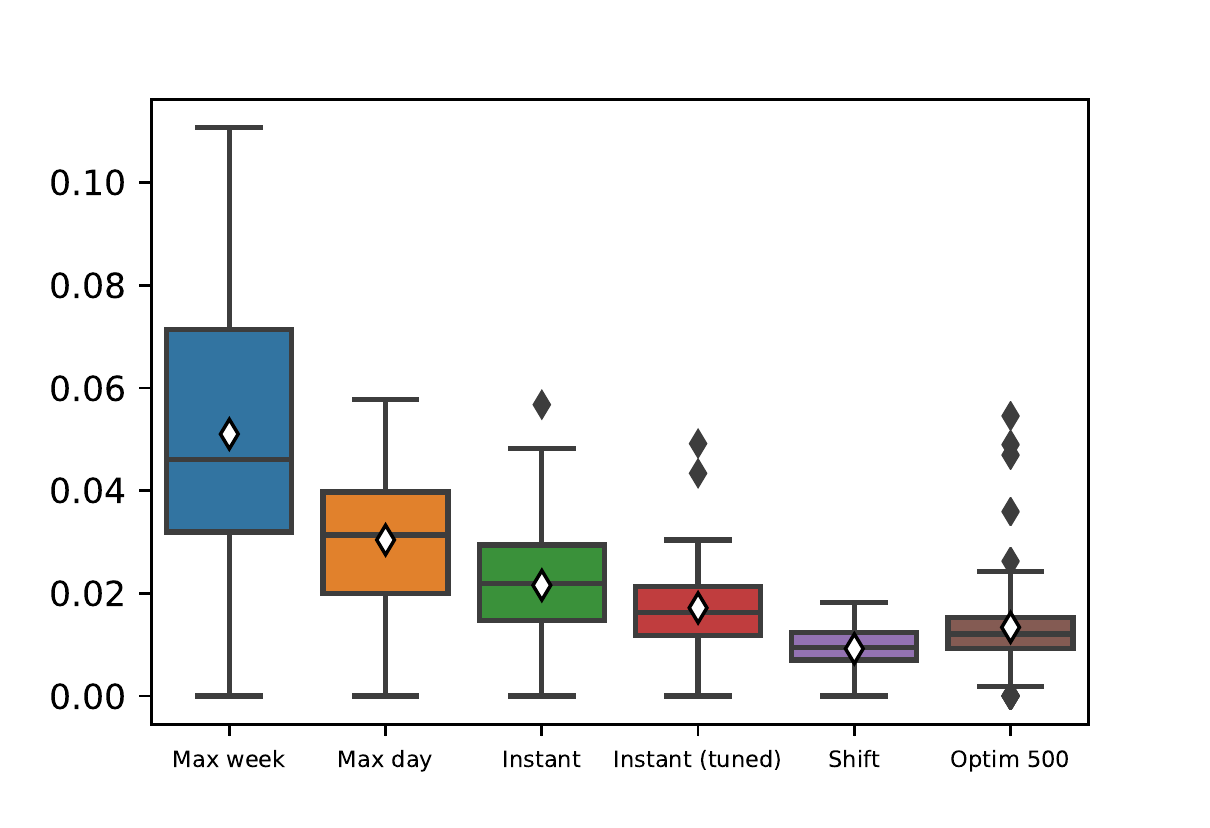}
      };
      \node[left=of img1, node distance=0cm, rotate=90, anchor=center,yshift=-1.2cm,font=\color{black}] {loss};
    \end{tikzpicture}
  \end{subfigure}\\[-3ex]
  \begin{subfigure}[t]{0.5\textwidth}
    \begin{tikzpicture}
	    \centering
      \node (img1) {
        \includegraphics[width=\linewidth, height=4cm]{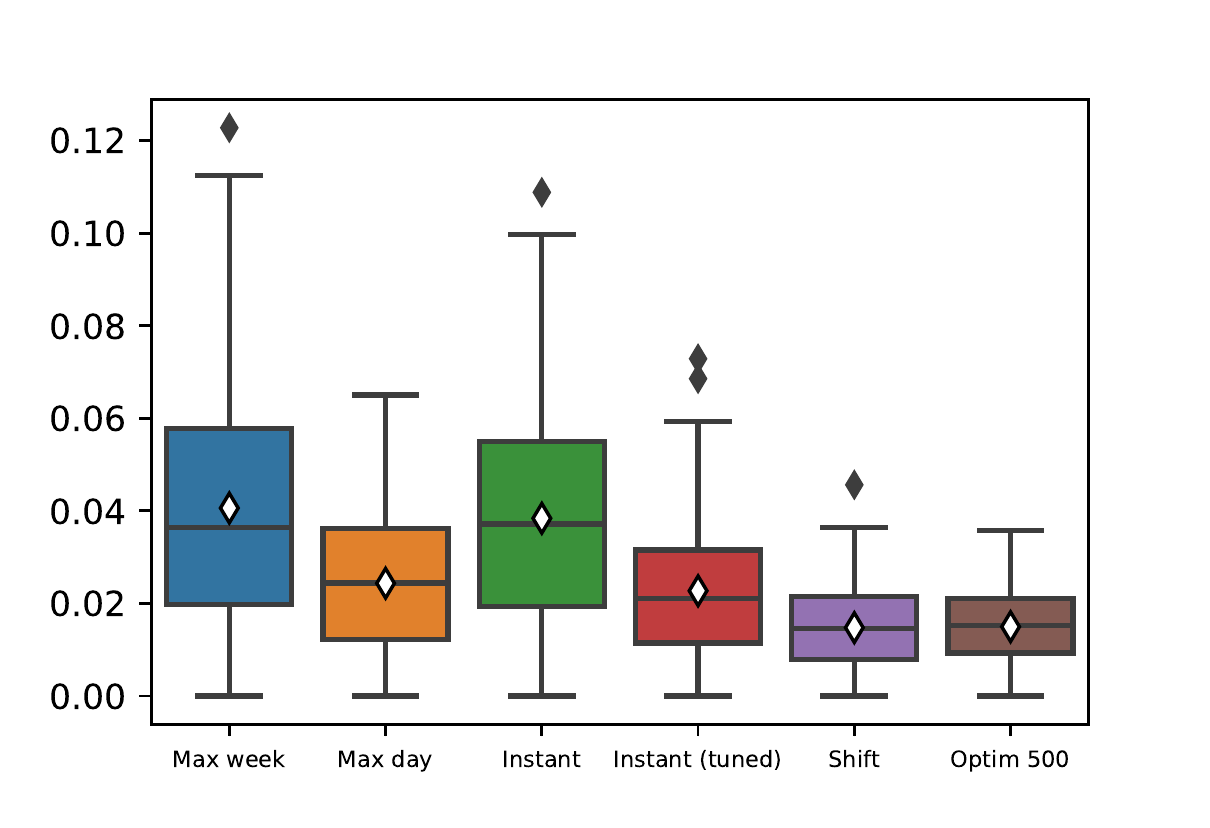}
      };
      \node[left=of img1, node distance=0cm, rotate=90, anchor=center,yshift=-1.2cm,font=\color{black}] {loss};
    \end{tikzpicture}
  \end{subfigure}\\[-3ex]
  \begin{subfigure}[t]{0.5\textwidth}
    \begin{tikzpicture}
	    \centering
      \node (img1) {
        \includegraphics[width=\linewidth, height=4cm]{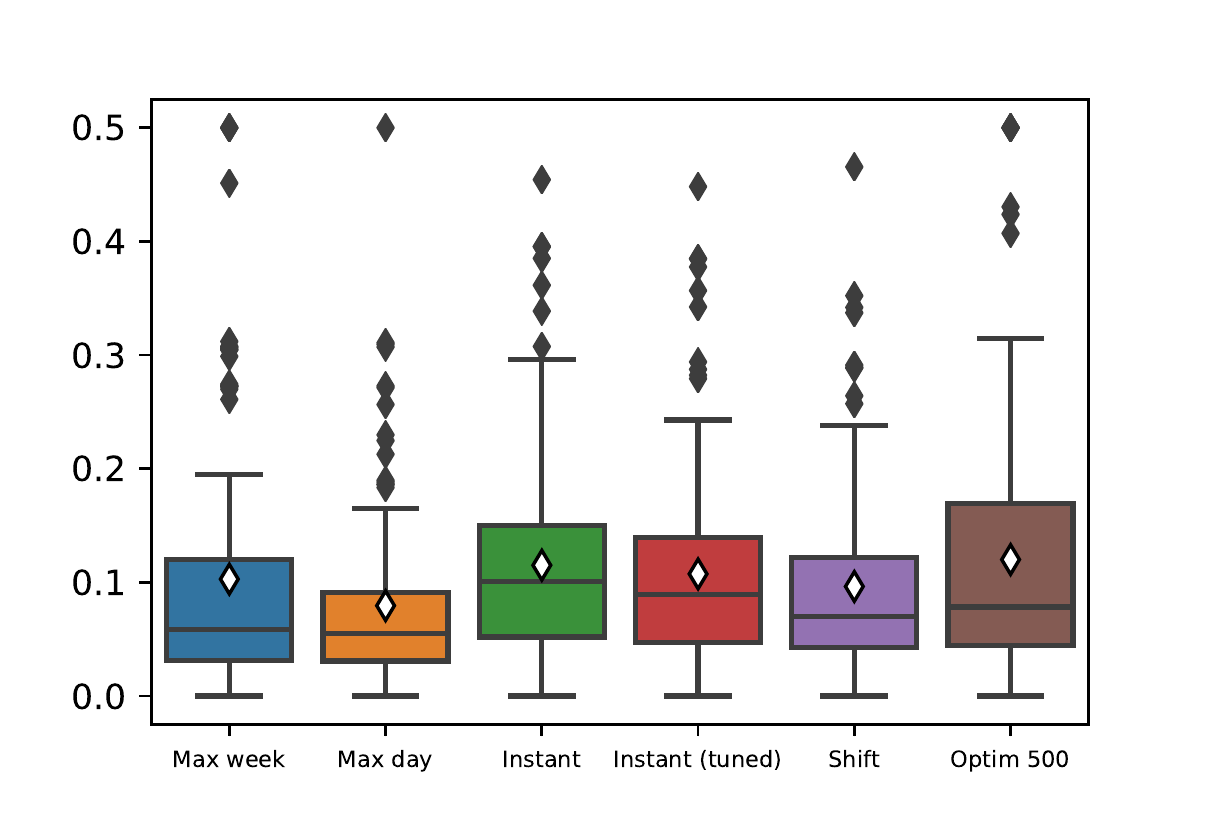}
      };
      \node[below=of img1, node distance=0cm, yshift=1.35cm,font=\color{black}] {policy};
      \node[left=of img1, node distance=0cm, rotate=90, anchor=center,yshift=-1.2cm,font=\color{black}] {loss};
    \end{tikzpicture}
  \end{subfigure}\\[-4ex]
	\caption{
    \label{fig:los_comparison}
    Losses obtained on the datasets $\boldsymbol{D}_1$ (top), $\boldsymbol{D}_2$
    (middle), and $\boldsymbol{R}_1$ (bottom) for the presented policies.
  }
\end{figure}
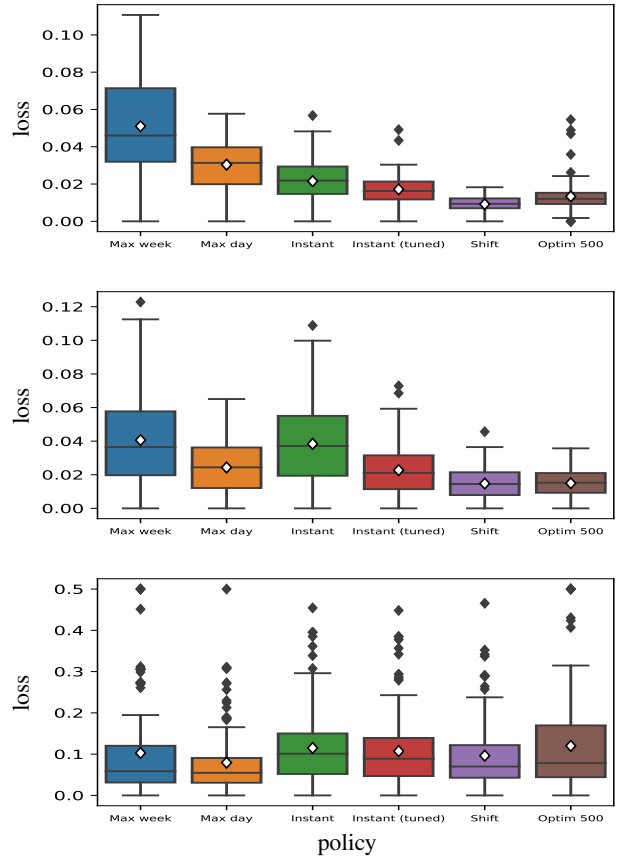

\header{Results} Figure~\ref{fig:los_comparison} provides the costs
associated to the scaling policies. 
The first two columns, \emph{Max week} and \emph{Max day} refer to the first heuristic,
where the maximum is taken over an entire week and a day respectively, leading to over-capacity.
\emph{Instant} and \emph{Instant (tuned)} refer to the reactive policy.
\emph{Shift} refers to the forecast shifting procedure depicted in
Algorithm~\ref{alg:forecast_shifting} and \emph{Optim 500} is the optimization
scaling policy for which the solver completes 500 iterations.

From all panels of Figures~\ref{fig:los_comparison} it is apparent that the
forecasting shifting policy compares favorably to other approaches in terms of
costs, in particular with respect to the optimal policies. This is particularly
visible on the two simulated datasets. On real-world data the median loss using
the shifting policy is larger than that of the Max day and Max week policies,
however the latter two result in larger extreme losses.  

Figure \ref{fig:qshift} shows the outcome of the optimization and the shift
policies on a time-series from $\boldsymbol{D}_1$.
It turns out that when $\alpha$ is close to 1, the scaling patterns of the
forecast shifting method converge to the ones obtained by optimizing the
objective function~\eqref{eq:optimization_problem}.
We also observe that the forecasting shifting policy comes at a fraction of the
optimization computational cost. Therefore, a practical predictive auto-scaling
solution should adopt this policy.
The datasets increase in strength of seasonality in relation to the noise level
as $\boldsymbol{R}_1 \to \boldsymbol{D}_2 \to \boldsymbol{D}_1$.
Intuitively, the stronger this signal to noise ratio is, the more important
seasonal scaling becomes.
For weak seasonality the simple max scaling baselines already work well -- no
scaling is necessary in the limit of very weak seasonality.
For large noise, the forecasting and optimization problems also become more
challenging, which increases the variance for the optimization method in this
case.

\section{Conclusion} \label{sec:conclusion}

We introduced a predictive auto-scaling problem formulation which takes a random
delay of resource availability and a limiting throughput into account.
Incorporating probabilistic forecasts, as opposed to merely point forecasts, we
obtain a rigorous mathematical optimization problem formulation for which we
provide a heuristics that works as well as optimal solutions in practically
relevant scenarios at a fraction of the compute cost. 
